\documentclass[]{iopart}

\usepackage{graphicx}	
\usepackage{color}
\usepackage{units}
\usepackage[separate-uncertainty=true, multi-part-units = single, product-units=single]{siunitx}
\graphicspath{ {Images/} }

\begin{document}

\title[]{Absolute absorption on the potassium D lines: theory and experiment}

\author{Ryan K Hanley, Philip D Gregory, Ifan G Hughes and Simon L Cornish}
\address{Joint Quantum Centre (JQC) Durham-Newcastle, Department of Physics, Durham University, South Road, Durham, DH1 3LE, United Kingdom}

\ead{ryan.hanley@durham.ac.uk}

\begin{abstract}
We present a detailed study of the absolute Doppler-broadened absorption of a probe beam scanned across the potassium D lines in a thermal vapour. Spectra using a weak probe were measured on the 4S $\rightarrow$ 4P transition and compared to the theoretical model of the electric susceptibility detailed by Zentile \etal (2015) in the code named ElecSus. Comparisons were also made on the 4S $\rightarrow$ 5P transition with an adapted version of ElecSus. This is the first experimental test of ElecSus on an atom with a ground state hyperfine splitting smaller than that of the Doppler width. An excellent agreement was found between ElecSus and experimental measurements at a variety of temperatures with rms errors $\sim 10^{-3}$. We have also demonstrated the use of ElecSus as an atomic vapour thermometry tool, and present a possible new measurement technique of transition decay rates which we predict to have a precision of $\sim$ $\SI{3}{kHz}$.
\end{abstract}

\maketitle


\section{Introduction}
The study of thermal vapours is a rapidly expanding field of physics with many applications. These range from all optical delay lines \cite{OpticalDelay}, quantum atomic memories \cite{lvovsky2009optical,sprague2014broadband, julsgaard2004experimental, quantumMem}, clocks \cite{knappe2004microfabricated, clock} and compact magnetometry \cite{subfemtotesla, budker2007optical, Magnetometer} to name but a few. The knowledge of the absorptive and dispersive properties of an atomic vapour has many benefits for existing and potential applications. This has been shown to facilitate the design of optical components, such as optical isolators \cite{OpticalIsolator} and atomic Farday filters \cite{lineBroadening,faraday,ZentileFaraday}, without the need for exhaustive trial and error. Furthermore, spectroscopy of alkali-metal elements, such as rubidium and caesium, is used for many laser-locking schemes \cite{LaserStab, davll,lecomte2000self}. Knowledge of the response of a thermal vapour to a weak resonant beam can aid the effective design of such systems. 

The absolute Doppler-broadened absorption of a probe beam resonant on the D lines of rubidium has been extensively studied by Siddons \etal \cite{Siddons}. They presented a theoretical framework which was utilised to correctly predict the absorption profile of a weak probe beam resonant on the D lines. Adapted from this theoretical framework, Zentile \etal developed `ElecSus' \cite{Elecsus}. ElecSus is a program which calculates the electric susceptibility of a thermal vapour in the vicinity of the alkali-metal D lines which in turn allows one to calculate many optical measurables, such as the transmission of a weak probe beam, the dispersive properties of the medium \cite{weller2012absolute} and the magnetic shifts of spectra \cite{zentile2014hyperfine}. The absolute Doppler-broadened absorption on the potassium D lines is yet to be studied extensively. The D lines of potassium are currently extensively used for gray molasses cooling \cite{salomon2013gray, gray,sievers2015simultaneous,nath2013quantum} to create ultracold ensembles of atoms, which are used to simulate many-body problems in physics \cite{bloch2008many, bloch2012quantum}. It has also been shown that using the 4S $\rightarrow$ 5P D line transitions allows for greater cooling of the potassium atoms due to the reduced linewidth \cite{KTrans5,mills2005lifetime}. Potassium is of particular interest due to the existence of both bosonic and fermionic species \cite{NIST, K39Spin,K41Spin,K40Spin} which allows one to study interactions in Bose-Fermi mixtures \cite{KBosonFermion}. 

Spectroscopy of potassium is different from that of rubidium due to the small spacing of its hyperfine energy levels. In particular, the ground state hyperfine splitting in potassium is less than the Doppler width and hence only one absorption profile is seen on each of the D lines \cite{bruner1998frequency,gustafsson2000atomic}. The aim of this work is to establish the weak-probe regime for the 4S $\rightarrow$ 4P and 4S $\rightarrow$ 5P D lines in potassium in order to test the quality of ElecSus on an atom where the ground state hyperfine splitting is less than the Doppler width.

The paper is structured as follows. In section 2, we give a brief description of the weak-probe regime and the physics behind ElecSus. Section 3 provides details of the experimental apparatus and procedure. Section 4 details the atomic structure of the potassium D line transitions along with our experimental results. Section 5 outlines a possible new experiment technique to measure transition decay rates. We conclude our findings in section 6. 

\section{Theory}
\subsection{Weak-probe regime}
The weak-probe regime of a resonant probe beam is defined as the incident intensity $I_0$ at which a the probe beam intensity is sufficiently weak that the absorption coefficient is independent of incident intensity. This regime in rubidium is typically several orders of magnitude smaller than the saturation intensity $I_{\rm{sat}}$ of the transition \cite{Siddons}. The saturation intensity of a two-level system is defined by 
\begin{equation} \label{eq:Isat}
I_{\rm{sat}} = \frac{\pi h c \Gamma }{3 \lambda ^3}~,
\end{equation}
where $h$ is Planck's constant, $c$ is the speed of light, $\Gamma$ is the linewidth and $\lambda$ is the wavelength of the transition \cite{Foot}.

Despite their simple atomic structure, alkali-metal atoms cannot be described by a simple two-level model. In particular, the presence of multiple hyperfine levels in the ground state can significantly alter the transmission of resonant probe light. Consider a system with two ground states $\left|g_{\rm{1}}\right\rangle$ and $\left|g_{\rm{2}}\right\rangle$ and one excited state $\left|e\right\rangle$ where the transitions $\left|g_{\rm{1}}\right\rangle \rightarrow \left|e\right\rangle$ and $\left|g_{\rm{2}}\right\rangle \rightarrow \left|e\right\rangle$ are allowed and $\left|g_{\rm{1}}\right\rangle \rightarrow \left|g_{\rm{2}}\right\rangle$ is forbidden. If the probe beam is resonant with the  $\left|g_{\rm{1}}\right\rangle \rightarrow \left|e\right\rangle$ transition, atoms are excited to $\left|e\right\rangle$ but they may spontaneously decay from $\left|e\right\rangle \rightarrow \left|g_{\rm{1}}\right\rangle$ or ${\left|e\right\rangle \rightarrow \left|g_{\rm{2}}\right\rangle}$. As the light is off-resonance with the ${\left|g_{\rm{2}}\right\rangle \rightarrow \left|e\right\rangle}$ transition, the atoms which have decayed into $\left|g_{\rm{2}}\right\rangle$ remain there. This leads to a depletion in the ground state population of ${\left|g_{\rm{1}}\right\rangle}$ and hence an increase in transmission of the probe beam. Redistributing the ground state populations using light is known as optical pumping \cite{Sherlock}. This system therefore has non-trivial dependencies on the determination of the weak-probe regime due to optical pumping.

\subsection{ElecSus}
The frequency dependent complex electric susceptibility of a dielectric $\chi \left(\Delta\right)$, where $\Delta$ is the detuning from resonance, connects the macroscopic polarisation to the applied electric field \cite{fox}. The real part $\chi _{\rm{R}} \left(\Delta\right)$ describes the dispersive properties and the imaginary part $\chi _{\rm{I}} \left(\Delta\right)$ describes the absorptive properties.

In order to calculate $\chi \left(\Delta\right)$ in the vicinity of the D line transitions, ElecSus first calculates the transition strengths of allowed electric-dipole transitions between all hyperfine sub-levels. To account for Doppler broadening due to the line-of-sight thermal motion of the atoms, a lineshape of the resonance is applied to each transition. This lineshape is a convolution of the Lorentzian atomic lineshape, dependent on $\Gamma$, and the Gaussian velocity distribution which is a function of temperature. Finally, $\chi \left(\Delta\right)$ is calculated by summing over all transitions. From $\chi \left(\Delta\right)$, many optical observables may be predicted.

As $\chi _{\rm{I}} \left(\Delta\right)$ describes the absorptive properties of the medium, the absorption coefficient $\alpha$ is given by $\alpha \left(\Delta\right) = k \chi _{\rm{I}} \left(\Delta\right)$ \cite{Siddons} where $k$ is the wavevector of the incident radiation. The absorption coefficient is dependent on the linewidth of the transition and the atomic number density ${\cal{N}}$ in the vapour which is exponential in the temperature, $T$. For a particular hyperfine transition, the absorption coefficient is given by \cite{Siddons}
\begin{equation} \label{eq:absorp}
\alpha\left(\Delta\right) = k C_{\rm{F}}^2 d^2 {\cal{N}} \frac{1}{2\left(2{\cal{I}}+1\right)} \frac{1}{\hbar \epsilon _0} s^{\rm{I}}\left(\Delta\right)~,
\end{equation}
where $ C_{\rm{F}}$ is the total transition strength, ${\cal{I}}$ is the nuclear spin, $\hbar$ is the reduced Planck's constant, $\epsilon_0$ is the permittivity, $s^{\rm{I}}\left(\Delta\right)$ is the transition lineshape and $d$ is given by
\begin{equation} \label{eq:Strength}
d=\sqrt{3}\sqrt{\frac{3 \epsilon_0 \hbar \Gamma \lambda^3}{8 \pi ^2}}~,
\end{equation}
where $\lambda$ is the wavelength of the transition. Therefore, using the Beer-Lambert law, one is able to predict the transmission of a weak probe beam resonant on the alkali-metal D lines. For rubidium and caesium a vapour cell of a few cm length at room temperature has sufficient number density to observe significant absorption features. However, due to the higher melting point of potassium \cite{Melting}, the vapour cell must be heated to ensure a large enough atomic vapour number density such that absorption features are observed. 

\section{Experimental details}

\begin{figure}
\includegraphics[width=\textwidth,angle=0]{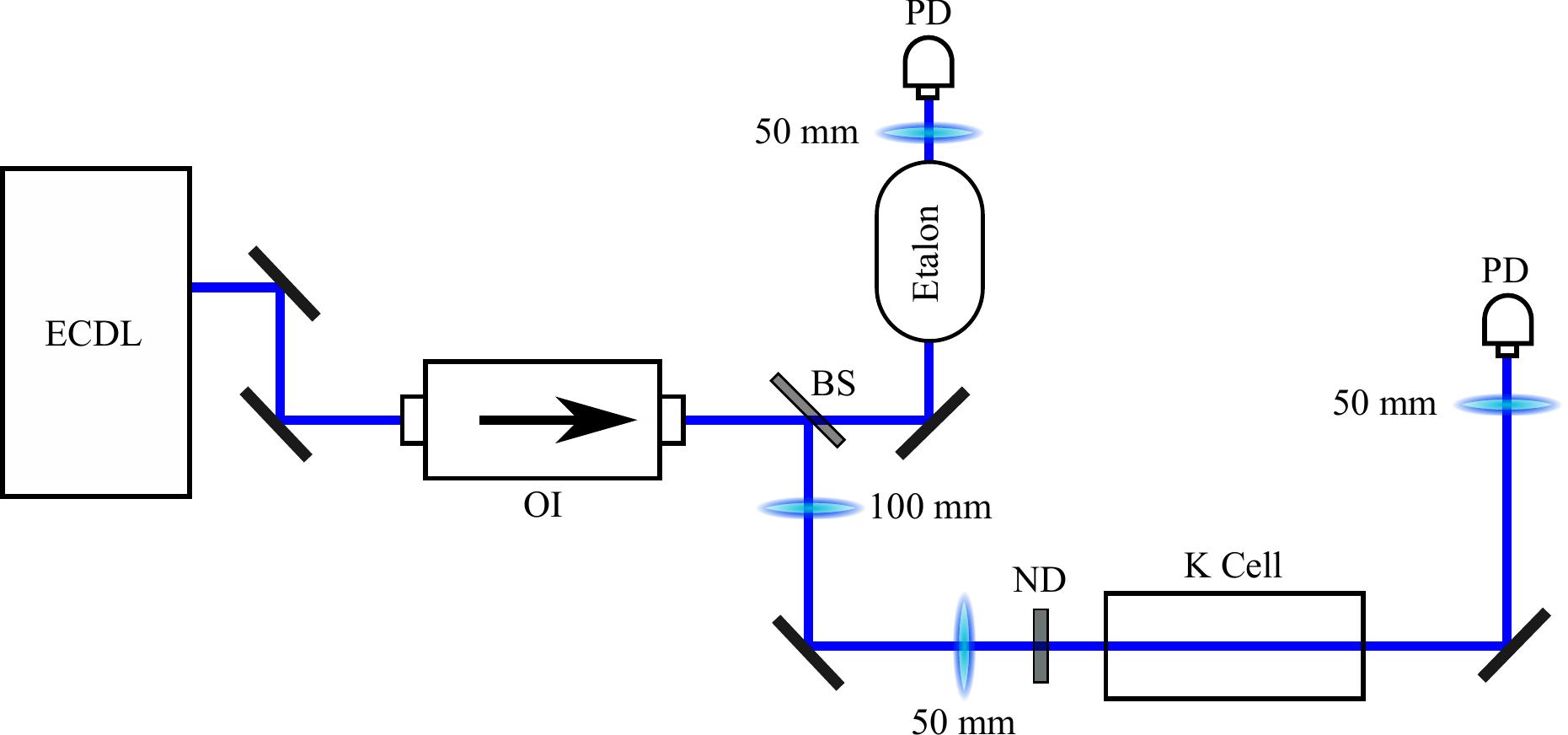}
\caption{Experimental configuration for the potassium spectroscopy. ECDL - extended cavity diode laser; OI - optical isolator; BS - 70:30 beam splitter; ND - neutral density filter; PD - photodiode; K Cell - heated potassium vapour cell (see main text). The $\SI{100}{mm}$ and $\SI{50}{mm}$ refer to the focal lengths of the lenses.}
\label{fig:setup}
\end{figure}

The experimental configuration is shown in Figure \ref{fig:setup}. A home-built external cavity diode laser (ECDL) in the Littrow configuration \cite{hawthorn2001littrow} was the light source. An Eagleyard EYP-RWE-0790-04000-0750-SOT01-0000 laser diode was used for the 4S $\rightarrow$ 4P transition and a Sanyo DL5146-101S laser diode was used for the 4S $\rightarrow$ 5P transition. The output beam from the laser passed through an optical isolator (OI) before impinging on a 70:30 beamsplitter (BS). The stronger beam was directed into a scanning Fabry-Perot etalon, which was used for frequency calibration. A Toptica FPI-100-0750-y scanning Fabry-Perot etalon with a free spectral range of $\SI{1.00 \pm 0.01}{GHz}$ and a Thorlabs SA200-3B scanning Fabry-Perot etalon with a free spectral range of $\SI{1.50 \pm 0.01}{GHz}$ were used for the 4S $\rightarrow$ 4P and 4S $\rightarrow$ 5P transitions respectively. The weaker beam was sent through a variable neutral density filter (ND), to ensure control of the probe beam intensity, before passing through a resistively heated, $\SI{10}{cm}$, natural abundance potassium vapour cell (K Cell). The magnetic field over the cell was measured to be of the order of $\SI{e-1}{G}$. Note that no attempt was made to null the field. The probe beam was then focused onto a calibrated photodiode (PD) which was integrated into a low-pass filter circuit in order to remove high-frequency noise and allow for greater sensitivity. The minimum detection power was $\sim \SI{1}{nW}$. 

The wavelength of light from the ECDL was tuned to the D lines by adjusting the angle of the diffraction grating. The wavelength was measured using a HighFinesse WS-5 wavelength meter to ensure the correct transitions were being probed. The laser was then scanned by applying a triangular ramp voltage at $\SI{10}{Hz}$ to a piezo stack actuator controlling the grating position. This allowed the laser to be scanned approximately $\SI{8}{GHz}$ before a mode hop. This was sufficiently large to scan the whole Doppler-broadened transition. Note that current feed-forward was employed on the blue 4S $\rightarrow$ 5P transition in order to increase the scan range of the laser.

The voltage output from the calibrated photodiode was recorded on a digital oscilloscope in order to record the transmission spectrum. The ramp voltage and etalon transmission peaks were also recorded. The etalon transmission peaks were used to convert from the measurement time into frequency, as well as linearising the scan. The spacing between the etalon peaks was plotted as a function of observed time. A polynomial was then fitted to a plot of the difference between observed and expected transmission peak times and subsequently used to linearise the scan.  

The weak-probe regime was established by varying the intensity of the probe beam incident on the cell from ${1\times 10^{-3}~I_{\text{sat}} \rightarrow 1\times 10^{2}~I_{\text{sat}}}$ by inserting a variety of neutral density filters into the path of the probe beam and measuring the minimum transmission. This was taken to be the mean over several points at the minimum of the transmission spectrum. Using a CCD camera, the $1/e^2$ widths of the probe beam were measured by taking cross sectional images of the probe beam before the cell and fitting Gaussian profiles to the images. The incident power was measured using the photodiode in the wings of the transmission spectrum. Spectra in the weak-probe regime were recorded at a variety of temperatures. Theoretical spectra generated by ElecSus were fit to the experimental spectra. A linear fit was made to the wings of the spectrum in order normalise the data to account for a weak variation in the laser power during the laser scan.

\section{Results and Discussion}
\subsection{4S $\rightarrow$ 4P D lines} 

\begin{figure}
\includegraphics[width=\textwidth,angle=0]{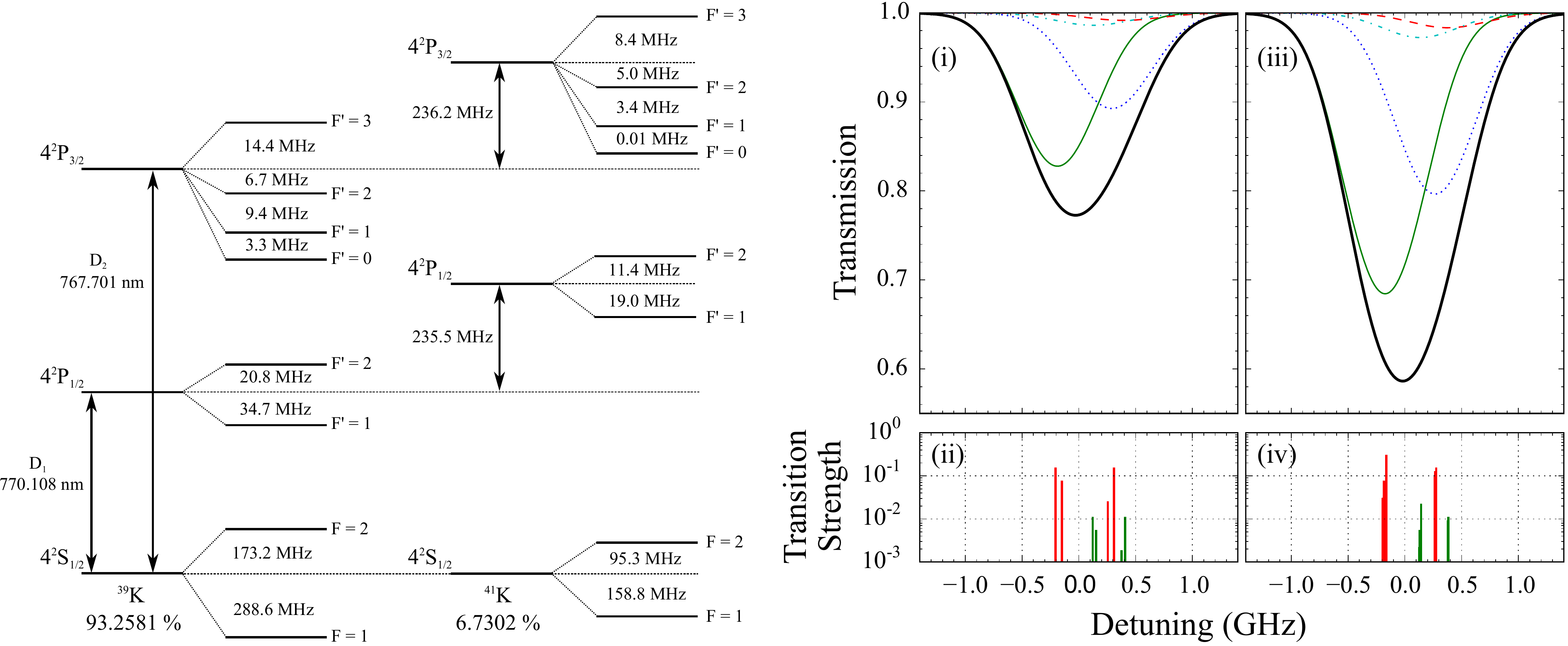}
\caption{An energy level diagram of the 4S $\rightarrow$ 4P D lines of $^{39}$K and $^{41}$K along with their natural abundance. Note that we have neglected $^{40}$K due to its natural abundance of $\SI{0.0117}{\%}$. Data taken from \cite{NIST,KTrans1,KTrans2,KTrans3,KTrans4,KTrans6}. Plots (i) and (iii) show the calculated contributions to the total transmission spectrum of a probe beam resonant with the D1 and D2 transition passing through a $\SI{10}{cm}$ natural abundance potassium vapour cell at $\SI{40}{\celsius}$ respectively. The blue (dotted) and green (solid) spectra correspond to transitions from $F=1 \rightarrow F^{\prime}$ and $F=2 \rightarrow F^{\prime}$ for $^{39}$K respectively. The red (dashed) and cyan (dot-dashed) spectra correspond to transitions from $F=1 \rightarrow F^{\prime}$ and $F=2 \rightarrow F^{\prime}$ for $^{41}$K respectively. The thick black profile is the total transmission spectrum. Plots (ii) and (iv) show the transition strengths, weighted by the isotopic abundance, where the red and green lines correspond to $^{39}$K and $^{41}$K respectively. The linear detuning ($\Delta/2\pi$) is referenced to the weighted line-centre of the transition.}
\label{fig:KStructure}
\end{figure}

In order to have measurable absorption of a probe beam with a good signal-to-noise ratio, the potassium vapour cell was heated to a typical temperature of $\sim \SI{40}{\celsius}$, as at this temperature there is sufficient atomic vapour number density to absorb a significant proportion of the incident beam. At a temperature of $\SI{40}{\celsius}$, the Doppler width is $\sim \SI{0.8}{GHz}$. This is much larger than the ground state hyperfine splitting of the potassium isotopes shown in Figure \ref{fig:KStructure}. Consequently, one would expect to observe only a single Doppler-broadened transmission spectrum for each of the D lines with contributions from both ground states. This can be seen in plots (i) and (iii) of Figure \ref{fig:KStructure}. The saturation intensity and decay rates for the D1 and D2 transitions are $\SI{1.71}{mW/cm^2}$, $\SI{1.75}{mW/cm^2}$, ${\Gamma/2\pi = \SI{5.956}{MHz}}$ and ${\Gamma/2\pi = \SI{6.035}{MHz}}$ respectively \cite{KLinewidths}.

\begin{figure} \label{Figure3}
\includegraphics[width=10cm,angle=0]{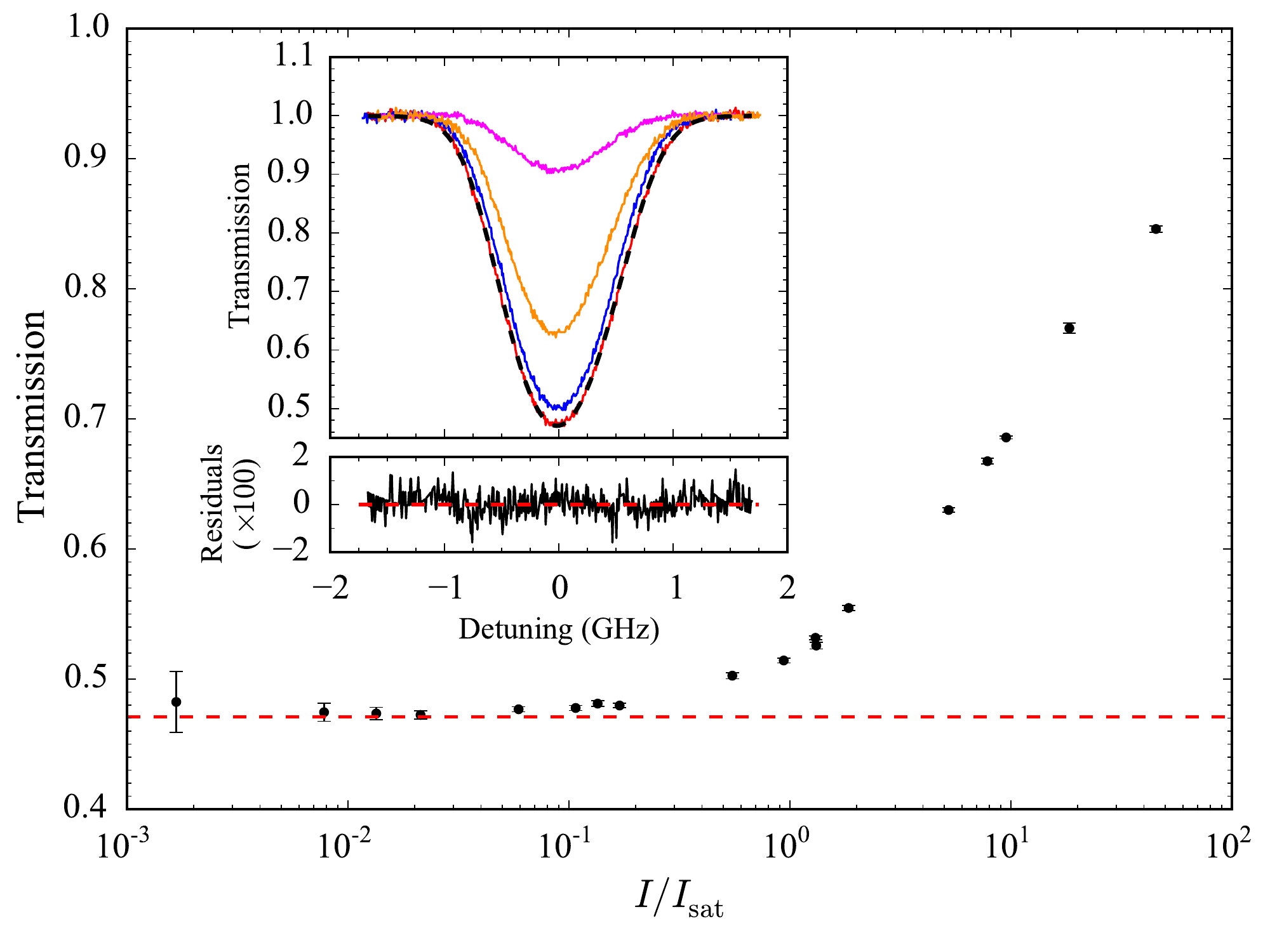}
\caption{The main figure shows a plot of the minimum transmission as a function of probe intensity, normalised to the saturation intensity, for a natural abundance potassium cell with light resonant on the D2 transition. The data corresponds to ${T=\SI{43.3 \pm 0.2}{\celsius}}$ and probe beam widths ${w_x = \SI{0.341 \pm 0.001}{mm}}$ and ${w_y = \SI{0.351 \pm 0.002}{mm}}$. The red dashed line is the theoretical value predicted by ElecSus. The inset is a plot of the transmission profile as a function of detuning from line centre, along with a theoretical fit (black dashed line). The four colours represent different probe intensities where $I_{\text{red}},~I_{\text{blue}},~I_{\text{orange}}~\text{and}~I_{\text{pink}}$ correspond to incident intensities of $\sim\left(\SI{8 e -3}{},~\SI{5 e -1}{},~\SI{4}{}~\text{and}~\SI{100}{}\right)$ $I_{\text{sat}}$ respectively. The residuals for the weak-probe regime are shown below.}
\label{fig:WeakProbe}
\end{figure}

In order to test ElecSus, the weak-probe regime was experimentally determined. In Figure \ref{fig:WeakProbe}, the main figure shows the minimum transmission of a probe beam as a function of incident intensity, normalised to the saturation intensity, of the D2 transition. The red data points show the minimum transmission of a probe beam, with $1/e^2$ widths of ${w_x = \SI{0.341 \pm 0.001}{mm}}$ and ${w_y = \SI{0.351 \pm 0.002}{mm}}$, through the vapour cell at a temperature of ${T=\SI{45 \pm 1}{\celsius}}$. The temperature was measured by inserting a thermocouple between the vapour cell and the surrounding metal heater assembly. The results demonstrate the significance of optical pumping. At low intensities, the probe beam is weak enough such that optical pumping effects are negligible and hence a change in probe beam intensity does not change the transmission. However, even at intensities $\sim I_{\rm{sat}}$ the scattering rate is sufficient to lead to a noticeable modification of the transmission. It is striking that at high intensities the amount of absorption is negligible compared to that of the weak-probe regime. From the transmission data, we experimentally established the weak-probe regime to be at an incident intensity of $\sim 10^{-1} ~I/I_{\text{sat}}$. The larger error bars at the low incident intensities are a consequence of low light levels and hence a poor signal-to-noise ratio.

The red data points in the inset of Figure \ref{Figure3} show a transmission spectrum in the weak-probe regime. The theory (blacked dashed line) is fitted to the data along with the transmission spectra for probe intensities higher than that of the weak probe to demonstrate the effect of optical pumping. The four transmission spectra colours represent the different probe intensities where $I_{\text{red}},~I_{\text{blue}},~I_{\text{orange}}~\text{and}~I_{\text{pink}}$ correspond to incident intensities of $\sim\left(\SI{8 e -3}{},~\SI{5 e -1}{},~\SI{4}{}~\text{and}~\SI{100}{}\right)$ $I_{\text{sat}}$ respectively. The residuals for the theoretical fit to the red data show limited structure and hence the quality of the theoretical model.

\begin{figure}
\includegraphics[width=\textwidth,angle=0]{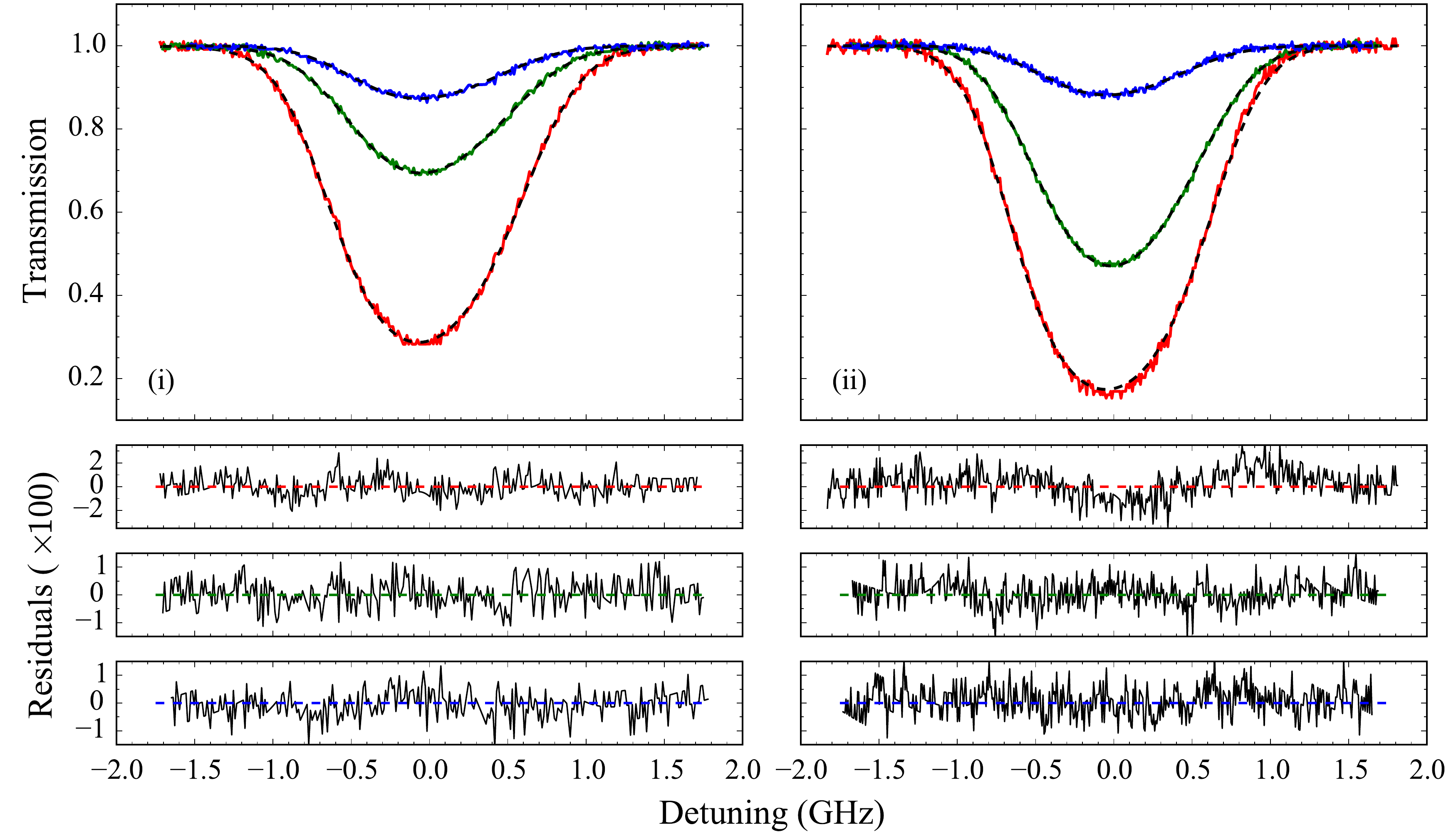}
\caption{The main figure shows a plot of the transmission as a function of temperature for a natural abundance potassium cell with a probe intensity in the weak-probe regime with light resonant on the D1 (i) and D2 (ii) transitions. The black dashed lines are the theoretical spectra predicted by ElecSus. The blue, green and red spectra correspond to temperatures of $\SI{34.0 \pm 0.7}{\celsius}$, $\SI{43.8 \pm 0.4}{\celsius}$ and $\SI{56.0 \pm 0.2}{\celsius}$ with rms errors between theory and experiment of $\SI{5 e -3}{}$, $\SI{5 e -3}{}$ and $\SI{8 e -3}{}$ respectively for (i) and $\SI{26.8 \pm 0.7}{\celsius}$, $\SI{43.3 \pm 0.5}{\celsius}$ and $\SI{51.7 \pm 0.4}{\celsius}$ with rms errors between theory and experiment of $\SI{5 e -3}{}$, $\SI{2 e -2}{}$ and $\SI{1 e -2}{}$ respectively for (ii). The residuals for each temperature are shown below.}
\label{fig:TFT1}
\end{figure}

A weak probe on the D1 and D2 transitions was passed through the cell at a variety of temperatures as shown in Figure \ref{fig:TFT1}, in order to further test the predictions of ElecSus. Plot (i) shows transmission spectra on the D1 transition at temperatures of $\SI{34.0 \pm 0.7}{\celsius}$, $\SI{43.8 \pm 0.4}{\celsius}$ and $\SI{56.0 \pm 0.2}{\celsius}$, as shown by the blue, green and red data points, with rms errors between theory and experiment of $\SI{5 e -3}{}$, $\SI{5 e -3}{}$ and $\SI{8 e -3}{}$ respectively. Here the temperature of the vapour was measured by fitting ElecSus to a weak-probe transmission spectrum, using a Marquardt-Levenberg algorithm to minimise the weighted squares of the residuals \cite{hughes2010measurements}. The errors on the temperatures were determined by calculating the $\chi^2_\nu +1$ statistic from the fits \cite{hughes2010measurements}. The fitted temperatures compare favourably with the thermocouple measurements, being consistently $\sim \SI{3}{\celsius}$ higher due to the fact that the thermocouple was poorly contacted to the cell. Plot (ii) shows transmission spectra on the D2 transition at temperatures of $\SI{26.8 \pm 0.7}{\celsius}$, $\SI{43.3 \pm 0.5}{\celsius}$ and $\SI{51.7 \pm 0.4}{\celsius}$ with rms errors between theory and experiment of $\SI{5 e -3}{}$, $\SI{2 e -3}{}$ and $\SI{1 e -2}{}$ respectively. The residuals for each of the three temperatures are shown below the main figures. All of the fits have a low rms errors and the residuals show limited structure which highlights the quality of the fit. The residuals for the red data points in (ii) show some undulating structure. This is believed to be due to an inadequate linearisation of the frequency scan as a consequence of a small scan range producing a limited number of etalon peaks. This therefore shows that the fitting procedure is sensitive to the scan linearisation. In rubidium for example, sub-Doppler spectroscopy is used to create many frequency markers to linearise the scan \cite{Siddons}. However, sub-Doppler spectroscopy of potassium does not lead to clear frequency markers due to the small hyperfine spacing \cite{pahwa2012polarization,mudarikwa2012sub} and hence this method was not employed. However, our method of linearisation appears to be sufficiently accurate, as shown by the quality of the other fits. 

\subsection{4S $\rightarrow$ 5P D lines} 

\begin{figure}
\includegraphics[width=\textwidth,angle=0]{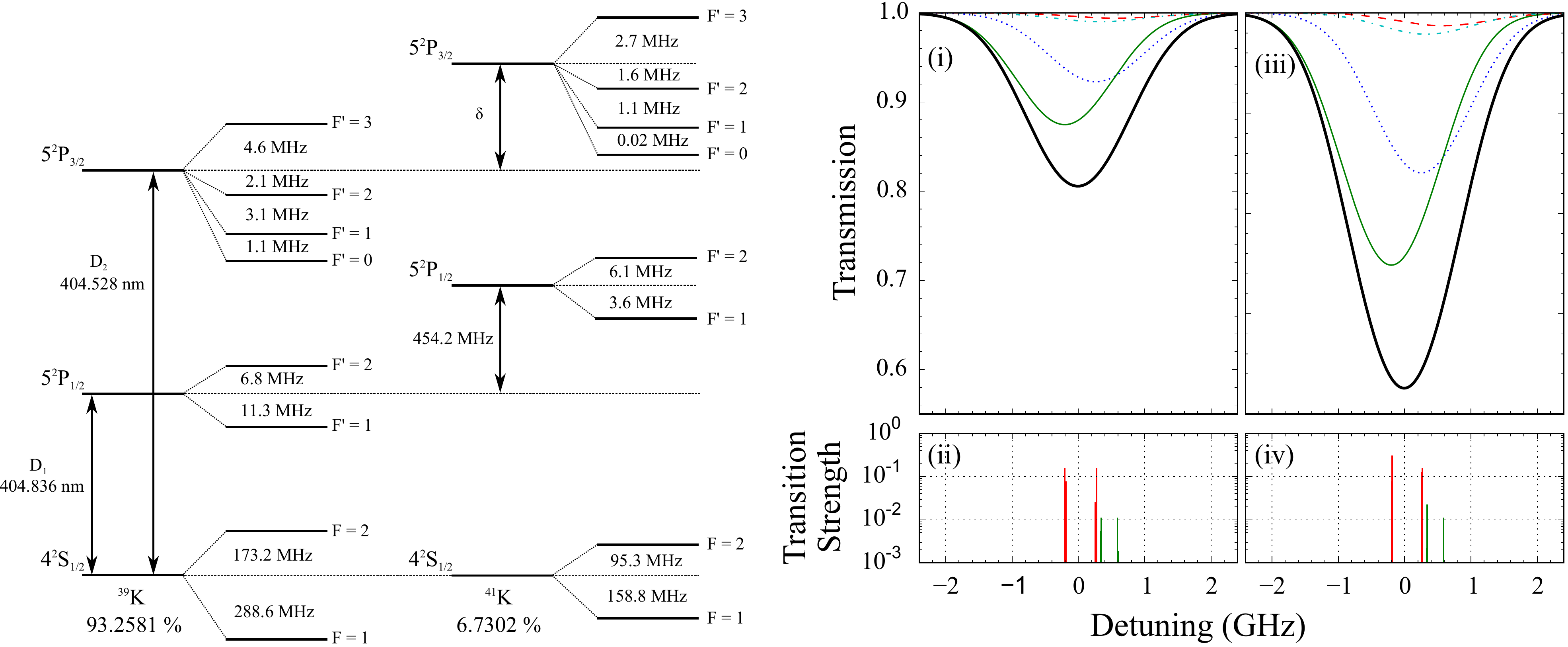}
\caption{An energy level diagram of the 4S $\rightarrow$ 5P D lines of $^{39}$K and $^{41}$K along with their natural abundance. Data taken from \cite{NIST,KTrans4,KTrans5,KTrans6,KTrans7,KTrans8,KTrans9}. Plots (i) and (iii) show the calculated contributions to the total transmission spectrum of a probe beam resonant with the D1 and D2 transition passing through a $\SI{10}{cm}$ natural abundance potassium vapour cell at $\SI{100}{\celsius}$ respectively. The colours correspond to those in Figure \ref{fig:KStructure}. Plots (ii) and (iv) show the transition strengths, weighted by the isotopic abundance, where the red and green lines correspond to $^{39}$K and $^{41}$K respectively. Detuning is reference to the weighted line-centre of the transition.}
\label{fig:4S5PStructure}
\end{figure}

The atomic structure of $^{39}$K and $^{41}$K on the 4S $\rightarrow$ 5P D lines is shown in Figure \ref{fig:4S5PStructure}. One should note that to the best of our knowledge, the isotope shift for the D2 transition has never been experimentally measured. Hence we have adopted the notation of $\delta$ to label the unknown isotope shift. The 4S $\rightarrow$ 5P transition is similar in structure to that of the 4S $\rightarrow$ 4P as an increase in principal quantum number does not change the angular components of the system. However, we note a different isotope shift and a smaller hyperfine splitting. The saturation intensity and decay rates for the D1 and D2 transitions are $\SI{58.8}{mW/cm^2}$, $\SI{57.6}{mW/cm^2}$, ${\Gamma/2\pi = \SI{170.3}{kHz}}$ and ${\Gamma/2\pi = \SI{184.6}{kHz}}$ respectively \cite{KTrans6}. One should note that equation \ref{eq:Isat} is not applicable here in calculating the saturation intensity as the 5P states may decay to alternative states. Therefore, all transitions must be taken into consideration \cite{5PData}. This leads to a saturation intensity of the 4S $\rightarrow$ 5P transition of 
\begin{equation}
I_{\rm{sat}}=2\pi \frac{h c}{\lambda ^3} \frac{\Gamma_{\rm{T}}^2}{\Gamma} \frac{1}{1.58}~,
\end{equation}
where $\Gamma_{\rm{T}}$ is the total linewidth of the transition. The factor of $1/1.58$ is a consequence of considering all decay paths from the 5P state.

ElecSus was adapted to model the ${4\text{S} \rightarrow 5\text{P}}$ transitions. This was easily implemented by changing the hyperfine constants, isotope shifts and decay rates. However, two decay rates were included in the code. $\Gamma$ was used to calculate the strength of the D line transitions, as shown in equation \ref{eq:Strength}, but the width of the excited state was controlled via $\Gamma_{\rm{T}}$. The total linewidth for the D1 and D2 transitions are $\Gamma_{\rm{T}}/2\pi = \SI{1.16}{MHz}$ \cite{mills2005lifetime} and $\Gamma_{\rm{T}}/2\pi = \SI{1.19}{MHz}$ \cite{KTrans5} respectively. As shown in equation \ref{eq:absorp}, the absorption coefficient is directly proportional to $\Gamma$ and ${\cal{N}}$. As $\Gamma$ for the 4S $\rightarrow$ 5P transition is much smaller than that of the 4S $\rightarrow$ 4P transition, in order to see a significant absorption feature, the vapour cell was heated to a typical temperature of $\SI{100}{\celsius}$ to increase ${\cal{N}}$. The transmission profile for a probe beam resonant with the D lines at a vapour temperature of $\SI{100}{\celsius}$ is shown in Figure \ref{fig:4S5PStructure}, which is similar to that of Figure \ref{fig:KStructure} and follows the same labelling conventions. We see a single Doppler-broadened spectrum with contributions from each hyperfine transition. However, the Doppler width is much larger, $\sim \SI{1.7}{GHz}$ compared to $\sim \SI{0.8}{GHz}$, as the Doppler width is directly proportional to the angular frequency of the transition.

Following the same method as before, the weak probe regime was experimentally determined to be $\sim \SI{1e-2}{} ~ I_\text{sat}$ for both the D1 and D2 transitions using probe beams of widths ${w_x = \SI{0.309 \pm 0.001}{mm}}$, ${w_y = \SI{0.907 \pm 0.008}{mm}}$ and $w_x = \SI{0.269 \pm 0.001}{mm}$, $w_y = \SI{0.623 \pm 0.006}{mm}$ at temperatures of $T = \SI{104 \pm 1}{\celsius}$ and $\SI{107 \pm 1}{\celsius}$ respectively where the temperatures were measured using the thermocouple.

\begin{figure} \label{Figure6}
\includegraphics[width=\textwidth,angle=0]{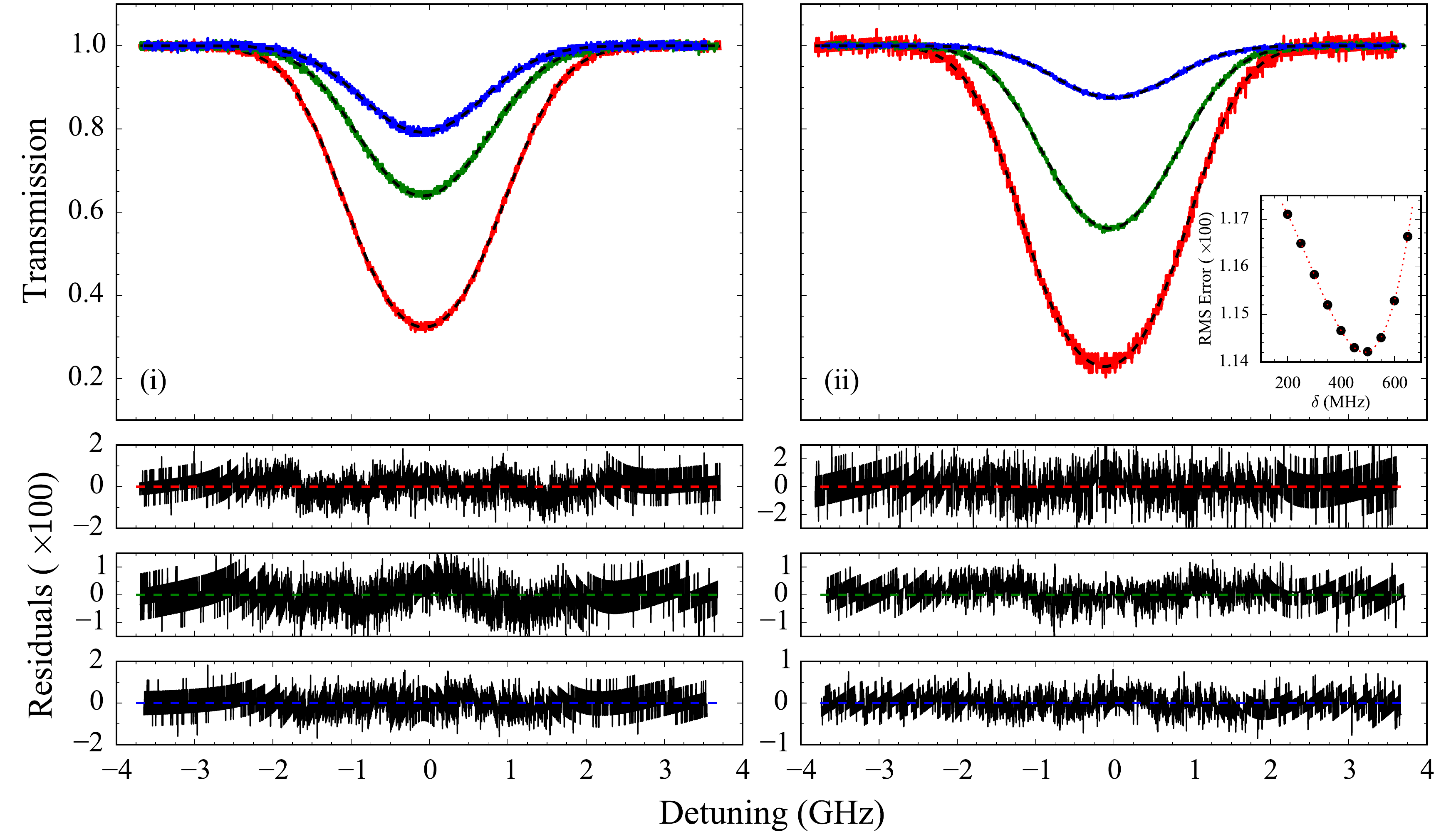}
\caption{The main figure shows a plot of the transmission as a function of temperature for a natural abundance potassium vapour cell with a probe intensity in the weak-probe regime with light resonant on the D1 (i) and D2 (ii) transitions. The black dashed lines are the theoretical spectra predicted by ElecSus. The blue, green and red spectra correspond to temperatures of $\SI{101.0 \pm 0.8}{\celsius}$, $\SI{110.7 \pm 0.4}{\celsius}$ and $\SI{125.4 \pm 0.3}{\celsius}$ with rms errors between theory and experiment of $\SI{5 e -3}{}$ for all three data sets for (i) and temperatures of $\SI{83.1 \pm 0.4}{\celsius}$, $\SI{103.0 \pm 0.2}{\celsius}$ and $\SI{117.0 \pm .7}{\celsius}$ with rms errors between theory and experiment of $\SI{2 e -3}{}$, $\SI{4 e -3}{}$ and $\SI{1 e -2}{}$ respectively for (ii). The residuals for each temperature are shown below. The inset shows the rms error between the red data and ElecSus as a function of the D2 isotope shift, $\delta$.}
\label{fig:TFT2}
\end{figure}

Figure \ref{fig:TFT2} shows the transmission of a weak probe on the D1 and D2 transitions at a variety of temperatures. Plot (i) shows transmission spectra on the D1 transition at temperatures of $\SI{101.0 \pm 0.8}{\celsius}$, $\SI{110.7 \pm 0.4}{\celsius}$ and $\SI{125.4 \pm 0.3}{\celsius}$, as shown by the blue, green and red data points, with rms errors between theory and experiment of $\SI{5 e -3}{}$ for all three data sets. Here the temperatures and their respective errors were measured using the fitting procedure detailed earlier. Plot (ii) shows transmission spectra on the D2 transition at temperatures of $\SI{83.1 \pm 0.4}{\celsius}$, $\SI{103.0 \pm 0.2}{\celsius}$ and $\SI{117.0 \pm .7}{\celsius}$ with rms errors between theory and experiment of $\SI{2 e -3}{}$, $\SI{4 e -3}{}$ and $\SI{1 e -2}{}$ respectively. One should note that the isotope shift used for the D2 transition was $\delta = \SI{454.2}{MHz}$ which is the isotope shift of the D1 transition. By considering Figure \ref{fig:KStructure}, we made the approximation that the difference between the isotope shifts of the D1 and the D2 transitions in Figure \ref{fig:4S5PStructure} would be of the order of $\SI{e0}{MHz}$. The small rms errors and the lack of structure in the residuals shows the quality of the agreement between theory and experimental data. Similar to that displayed in Figure \ref{fig:TFT1}, there is a small amount of undulation in some of the residuals. This is once again believed to be attributed to the quality of the linearisation of the laser scan. However, these data indicate that the theory and experiment are in agreement, considering the scale of the residuals and the rms errors. The level of noise appears to vary across different spectra. This is believed to be attributed to the scale on the oscilloscope at which the data was recorded. This is a significant result as it is indicative that the adapted version of ElecSus is a very accurate model of the 4S $\rightarrow$ 5P transition which could prove fruitful in future experiments.

The inset in Figure \ref{Figure6} plot (ii) shows the rms error between the red data and ElecSus as a function of the D2 isotope shift, $\delta$. This procedure was carried out in an attempt to measure $\delta$ as one would expect that the rms error would be minimised for the correct value of $\delta$. This procedure was also carried out on the other two fits in plot (ii), however the values for $\delta$ determined were not in agreement with each other. We believe this is a manifestation of poor scan linearisation. Although we are not able to identify $\delta$, this is significant as it does highlight that the quality of the agreement between the transmission spectra and ElecSus is sensitive to isotope shifts and hence, may be used to measure them. We believe that if one were to improve the signal-to-noise ratio and the scan linearity, it should be possible to measure $\delta$. One possibility for improving the scan linearisation would be to scan the probe light across the transition using a broadly tunable waveguide EOM driver by an rf synthesiser, avoiding the non-linearities associated with scanning the ECDL.

\section{Future Experiment}

The plots in Figure \ref{fig:TFT2} show that ElecSus can be adapted to accurately model the 4S $\rightarrow$ 5P transition. This leads one to consider an experiment where two parallel beams resonant with the 4S $\rightarrow$ 4P and 4S $\rightarrow$ 5P transitions propagate though the same vapour cell. The 4S $\rightarrow$ 4P beam may be used to accurately determine the temperature of the atoms to a precision of $\sim \SI{0.2}{\celsius}$. The other beam may be used for spectroscopy of the atoms. For example, the line centre transmission is dependent on the strength of the transition which is directly proportional to $\Gamma$. Therefore, having characterised the atoms with one probe beam, we would be able to fit to experimental data of the transmission of the other probe beam in order to measure the decay rate of the transition. This two beam method could also prove useful in experiments in which the temperature of an atomic vapour must be known accurately, such as experiments attempting to measure the Boltzmann constant from the Doppler-width of an absorption feature \cite{Boltz1,Boltz2}.

By varying $\Gamma$ for the 4S $\rightarrow$ 5P transition in simulations and assuming the best signal-to-noise we experimentally obtained in our experiment, we believe one should be able to determine $\Gamma$ to a precision of $\sim \SI{3}{kHz}$ at a temperature of $\SI{100}{\celsius}$ which is comparable to that given in the literature which is of the order of $\SI{}{kHz}$ \cite{KTrans6}. If the improvements to experimental noise were implemented, then it should be possible to surpass the experimental error on $\Gamma$ given in the literature. 

\section{Concluding Remarks}

Transmission spectra using a weak probe were measured on the 4S $\rightarrow$ 4P and 4S $\rightarrow$ 5P transitions. Theoretical spectra generated by ElecSus, and its adapted version, were fit to the experimental spectra. This is the first experimental test of ElecSus on an atom with a ground state hyperfine splitting smaller than that of the Doppler width. An excellent agreement was found between ElecSus and experimental measurements at a variety of temperatures on both transitions with rms errors $\sim\SI{e -3}{}$. This also shows that ElecSus may be adapted with relative ease to model higher principal quantum number transitions with similar accuracy. This validation of ElecSus on potassium will enable accurate design and implementation of future experiments where the absorptive and dispersive properties of a potassium vapour are required, such as optical isolators, Faraday filters and laser locking schemes. Furthermore, we have shown the sensitivity of the agreement between spectra and ElecSus to isotope shifts and proposed improvements to the current experiment which should facilitate more accurate measurements. We have also discussed the possibility of an experiment where two probe beams could be used to simultaneously interrogate an atomic vapour on different transitions. Using one beam for thermometry enhances the accuracy of the information which can be determined with the second beam, allowing accurate decay rate measurements for example.

\ack
We are grateful for the loan of equipment and discussion from Steve Hopkins and Stefan Kemp, as well as the detailed help with the core of ElecSus from Mark Zentile. We would also like to thank Michael K\"oppinger for the design of the potassium vapour cell heater. This work was financed by Durham University and EPSRC (grants EP/L023024/1 and EP/I012044/1). 

\section*{References}

\bibliographystyle{unsrt}
\bibliography{bibfile}

\begin{thebibliography}{10}

\bibitem{OpticalDelay}
Ryan~M Camacho, Michael~V Pack, John~C Howell, Aaron Schweinsberg, and Robert~W
  Boyd.
\newblock Wide-bandwidth, tunable, multiple-pulse-width optical delays using
  slow light in cesium vapor.
\newblock {\em Physical Review Letters}, 98(15):153601, 2007.

\bibitem{lvovsky2009optical}
Alexander~I Lvovsky, Barry~C Sanders, and Wolfgang Tittel.
\newblock Optical quantum memory.
\newblock {\em Nature photonics}, 3(12):706--714, 2009.

\bibitem{sprague2014broadband}
MR~Sprague, PS~Michelberger, TFM Champion, DG~England, J~Nunn, X-M Jin,
  WS~Kolthammer, A~Abdolvand, P~St~J Russell, and IA~Walmsley.
\newblock Broadband single-photon-level memory in a hollow-core photonic
  crystal fibre.
\newblock {\em Nature Photonics}, 8(4):287--291, 2014.

\bibitem{julsgaard2004experimental}
Brian Julsgaard, Jacob Sherson, J~Ignacio Cirac, Jarom{\'\i}r
  Fiur{\'a}{\v{s}}ek, and Eugene~S Polzik.
\newblock Experimental demonstration of quantum memory for light.
\newblock {\em Nature}, 432(7016):482--486, 2004.

\bibitem{quantumMem}
SD~Jenkins, YO~Dudin, R~Zhao, DN~Matsukevich, A~Kuzmich, and TAB Kennedy.
\newblock In situ determination of zeeman content of collective atomic
  memories.
\newblock {\em Journal of Physics B: Atomic, Molecular and Optical Physics},
  45(12):124006, 2012.

\bibitem{knappe2004microfabricated}
Svenja Knappe, Vishal Shah, Peter~DD Schwindt, Leo Hollberg, John Kitching,
  Li-Anne Liew, and John Moreland.
\newblock A microfabricated atomic clock.
\newblock {\em Applied Physics Letters}, 85(9):1460--1462, 2004.

\bibitem{clock}
James Camparo.
\newblock The rubidium atomic clock and basic research.
\newblock {\em Physics Today}, 60(11):33, 2007.

\bibitem{subfemtotesla}
IK~Kominis, TW~Kornack, JC~Allred, and MV~Romalis.
\newblock A subfemtotesla multichannel atomic magnetometer.
\newblock {\em Nature}, 422(6932):596--599, 2003.

\bibitem{budker2007optical}
Dmitry Budker and Michael Romalis.
\newblock Optical magnetometry.
\newblock {\em Nature Physics}, 3(4):227--234, 2007.

\bibitem{Magnetometer}
Peter~DD Schwindt, Svenja Knappe, Vishal Shah, Leo Hollberg, John Kitching,
  Li-Anne Liew, and John Moreland.
\newblock Chip-scale atomic magnetometer.
\newblock {\em Applied Physics Letters}, 85(26):6409--6411, 2004.

\bibitem{OpticalIsolator}
L~Weller, KS~Kleinbach, MA~Zentile, S~Knappe, IG~Hughes, and CS~Adams.
\newblock Optical isolator using an atomic vapor in the hyperfine paschen--back
  regime.
\newblock {\em Optics Letters}, 37(16):3405--3407, 2012.

\bibitem{lineBroadening}
Mark~A Zentile, Renju~S Mathew, Daniel~J Whiting, James Keaveney, Charles~S
  Adams, and Ifan~G Hughes.
\newblock Effect of line broadening on the performance of faraday filters.
\newblock {\em arXiv preprint arXiv:1504.03651}, 2015.

\bibitem{faraday}
Matthias Widmann, Simone Portalupi, Sang-Yun Lee, Peter Michler, J{\"o}rg
  Wrachtrup, and Ilja Gerhardt.
\newblock Faraday filtering on the cs-d $ \_1 $-line for quantum hybrid
  systems.
\newblock {\em arXiv preprint arXiv:1505.01719}, 2015.

\bibitem{ZentileFaraday}
Mark~A. Zentile, Daniel~J. Whiting, James Keaveney, Charles~S. Adams, and
  Ifan~G. Hughes.
\newblock Atomic faraday filter with equivalent noise bandwidth less than 1
  ghz.
\newblock {\em Opt. Lett.}, 40(9):2000--2003, 2015.

\bibitem{LaserStab}
Christoph Affolderbach and Gaetano Mileti.
\newblock A compact laser head with high-frequency stability for rb atomic
  clocks and optical instrumentation.
\newblock {\em Review of scientific instruments}, 76(7):073108, 2005.

\bibitem{davll}
Alfred Millett-Sikking, Ifan~G Hughes, Patrick Tierney, and Simon~L Cornish.
\newblock Davll lineshapes in atomic rubidium.
\newblock {\em Journal of Physics B: Atomic, Molecular and Optical Physics},
  40(1):187, 2007.

\bibitem{lecomte2000self}
Steve Lecomte, Emmanuel Fretel, Gaetano Mileti, and Pierre Thomann.
\newblock Self-aligned extended-cavity diode laser stabilized by the zeeman
  effect on the cesium d 2 line.
\newblock {\em Applied optics}, 39(9):1426--1429, 2000.

\bibitem{Siddons}
Paul Siddons, Charles~S Adams, Chang Ge, and Ifan~G Hughes.
\newblock Absolute absorption on rubidium {D} lines: comparison between theory
  and experiment.
\newblock {\em Journal of Physics B: Atomic, Molecular and Optical Physics},
  41(15):155004, 2008.

\bibitem{Elecsus}
Mark~A Zentile, James Keaveney, Lee Weller, Daniel~J Whiting, Charles~S Adams,
  and Ifan~G Hughes.
\newblock Elecsus: A program to calculate the electric susceptibility of an
  atomic ensemble.
\newblock {\em Computer Physics Communications}, 189:162--174, 2015.

\bibitem{weller2012absolute}
Lee Weller, Kathrin~S Kleinbach, Mark~A Zentile, Svenja Knappe, Charles~S
  Adams, and Ifan~G Hughes.
\newblock Absolute absorption and dispersion of a rubidium vapour in the
  hyperfine paschen--back regime.
\newblock {\em Journal of Physics B: Atomic, Molecular and Optical Physics},
  45(21):215005, 2012.

\bibitem{zentile2014hyperfine}
Mark~A Zentile, Rebecca Andrews, Lee Weller, Svenja Knappe, Charles~S Adams,
  and Ifan~G Hughes.
\newblock The hyperfine paschen--back faraday effect.
\newblock {\em Journal of Physics B: Atomic, Molecular and Optical Physics},
  47(7):075005, 2014.

\bibitem{salomon2013gray}
Guillaume Salomon, Lauriane Fouch{\'e}, Pengjun Wang, Alain Aspect, Philippe
  Bouyer, and Thomas Bourdel.
\newblock Gray-molasses cooling of 39k to a high phase-space density.
\newblock {\em EPL (Europhysics Letters)}, 104(6):63002, 2013.

\bibitem{gray}
D~Fernandes, F~Sievers1, N~Kretzschmar1, S~Wu, C~Salomon, and F~Chevy.
\newblock Sub-doppler laser cooling of fermionic 40k atoms in three-dimensional
  gray optical molasses.
\newblock {\em EPL (Europhysics Letters)}, 100:63001, 2012.

\bibitem{sievers2015simultaneous}
Franz Sievers, Norman Kretzschmar, Diogo~Rio Fernandes, Daniel Suchet, Michael
  Rabinovic, Saijun Wu, Colin~V Parker, Lev Khaykovich, Christophe Salomon, and
  Fr{\'e}d{\'e}ric Chevy.
\newblock Simultaneous sub-doppler laser cooling of fermionic li 6 and k 40 on
  the d 1 line: Theory and experiment.
\newblock {\em Physical Review A}, 91(2):023426, 2015.

\bibitem{nath2013quantum}
Dipankar Nath, R~Kollengode Easwaran, G~Rajalakshmi, and CS~Unnikrishnan.
\newblock Quantum-interference-enhanced deep sub-doppler cooling of 39 k atoms
  in gray molasses.
\newblock {\em Physical Review A}, 88(5):053407, 2013.

\bibitem{bloch2008many}
Immanuel Bloch, Jean Dalibard, and Wilhelm Zwerger.
\newblock Many-body physics with ultracold gases.
\newblock {\em Reviews of Modern Physics}, 80(3):885, 2008.

\bibitem{bloch2012quantum}
Immanuel Bloch, Jean Dalibard, and Sylvain Nascimb{\`e}ne.
\newblock Quantum simulations with ultracold quantum gases.
\newblock {\em Nature Physics}, 8(4):267--276, 2012.

\bibitem{KTrans5}
DC~McKay, Dylan Jervis, DJ~Fine, JW~Simpson-Porco, GJA Edge, and JH~Thywissen.
\newblock Low-temperature high-density magneto-optical trapping of potassium
  using the open 4 {S} $\rightarrow$ 5 {P} transition at 405 nm.
\newblock {\em Physical Review A}, 84(6), 2011.

\bibitem{mills2005lifetime}
A~Mills, JA~Behr, LA~Courneyea, and MR~Pearson.
\newblock Lifetime of the potassium 5 p 1/ 2 state.
\newblock {\em Physical Review A}, 72(2):024501, 2005.

\bibitem{NIST}
National {I}nstitute of {S}tandards and {T}echnology: {A}tomic {W}eights and
  {I}sotopic {C}ompositions.
\newblock http://physics.nist.gov/PhysRefData/Compositions/index.html, October
  2014.

\bibitem{K39Spin}
P.~Kusch, S.~Millman, and I.~I. Rabi.
\newblock The nuclear magnetic moments of {N}$^{14}$, {Na}$^{23}$, {K}$^{39}$
  and {Cs}$^{133}$.
\newblock {\em Phys. Rev.}, 55:1176--1181, 1939.

\bibitem{K41Spin}
J.~H. Manley.
\newblock The nuclear spin and magnetic moment of potassium (41).
\newblock {\em Phys. Rev.}, 49:921--924, 1936.

\bibitem{K40Spin}
J.~R. Zacharias.
\newblock The nuclear spin and magnetic moment of {K}$^{40}$.
\newblock {\em Phys. Rev.}, 61:270--276, 1942.

\bibitem{KBosonFermion}
Cheng-Hsun Wu, Ibon Santiago, Jee~Woo Park, Peyman Ahmadi, and Martin~W
  Zwierlein.
\newblock Strongly interacting isotopic bose-fermi mixture immersed in a fermi
  sea.
\newblock {\em Physical Review A}, 84(1):011601, 2011.

\bibitem{bruner1998frequency}
Ariel Bruner, Ady Arie, Mark~A Arbore, and Martin~M Fejer.
\newblock Frequency stabilization of a diode laser at 1540 nm by locking to
  sub-doppler lines of potassium at 770 nm.
\newblock {\em Applied optics}, 37(6):1049--1052, 1998.

\bibitem{gustafsson2000atomic}
U~Gustafsson, J~Alnis, and Sune Svanberg.
\newblock Atomic spectroscopy with violet laser diodes.
\newblock {\em American Journal of Physics}, 68(7):660--664, 2000.

\bibitem{Foot}
C.J. Foot.
\newblock {\em Atomic {P}hysics}.
\newblock Oxford University Press, 2005.

\bibitem{Sherlock}
Ben~E Sherlock and Ifan~G Hughes.
\newblock How weak is a weak probe in laser spectroscopy?
\newblock {\em American Journal of Physics}, 77(2):111--115, 2009.

\bibitem{fox}
M.~Fox.
\newblock {\em Optical Properties of Solids}.
\newblock OUP Oxford, 2010.

\bibitem{Melting}
CB~Alcock, VP~Itkin, and MK~Horrigan.
\newblock Vapour pressure equations for the metallic elements: 298--2500k.
\newblock {\em Canadian Metallurgical Quarterly}, 23(3):309--313, 1984.

\bibitem{hawthorn2001littrow}
CJ~Hawthorn, KP~Weber, and RE~Scholten.
\newblock Littrow configuration tunable external cavity diode laser with fixed
  direction output beam.
\newblock {\em Review of scientific instruments}, 72(12):4477--4479, 2001.

\bibitem{KLinewidths}
H~Wang, P~Gould, and W~Stwalley.
\newblock Long-range interaction of the 39k(4s)+ 39k(4p) asymptote by
  photoassociative spectroscopy. i. the 0−g pure long-range state and the
  long-range potential constants.
\newblock {\em Journal of Chemical Physics}, 106:7899, 1997.

\bibitem{KTrans1}
Stephan Falke, Eberhard Tiemann, Christian Lisdat, Harald Schnatz, and Gesine
  Grosche.
\newblock Transition frequencies of the {D} lines of {K} 39, {K} 40, and {K} 41
  measured with a femtosecond laser frequency comb.
\newblock {\em Physical Review A}, 74(3):032503, 2006.

\bibitem{KTrans2}
F.~Touchard, P.~Guimbal, S.~Büttgenbach, R.~Klapisch, M.~De~Saint Simon, J.M.
  Serre, C.~Thibault, H.T. Duong, P.~Juncar, S.~Liberman, J.~Pinard, and J.L.
  Vialle.
\newblock Isotope shifts and hyperfine structure of 38 - 47 {K} by laser
  spectroscopy.
\newblock {\em Physics Letters B}, 108(3):169 -- 171, 1982.

\bibitem{KTrans3}
N~Bendali, HT~Duong, and JL~Vialle.
\newblock High-resolution laser spectroscopy on the {D}1 and {D}2 lines of 39,
  40, 41 {K} using rf modulated laser light.
\newblock {\em Journal of Physics B: Atomic and Molecular Physics},
  14(22):4231, 1981.

\bibitem{KTrans4}
E~Arimondo, M~Inguscio, and P~Violino.
\newblock Experimental determinations of the hyperfine structure in the alkali
  atoms.
\newblock {\em Reviews of Modern Physics}, 49(1):31, 1977.

\bibitem{KTrans6}
JE~Sansonetti.
\newblock Wavelengths, transition probabilities, and energy levels for the
  spectra of potassium ({KI} through {KXIX}).
\newblock {\em Journal of Physical and Chemical Reference Data}, 37(1):7--96,
  2008.

\bibitem{hughes2010measurements}
I.~Hughes and T.~Hase.
\newblock {\em Measurements and their Uncertainties: A practical guide to
  modern error analysis}.
\newblock OUP Oxford, 2010.

\bibitem{pahwa2012polarization}
K~Pahwa, L~Mudarikwa, and J~Goldwin.
\newblock Polarization spectroscopy and magnetically-induced dichroism of the
  potassium d 2 lines.
\newblock {\em Optics express}, 20(16):17456--17466, 2012.

\bibitem{mudarikwa2012sub}
L~Mudarikwa, K~Pahwa, and J~Goldwin.
\newblock Sub-doppler modulation spectroscopy of potassium for laser
  stabilization.
\newblock {\em Journal of Physics B: Atomic, Molecular and Optical Physics},
  45(6):065002, 2012.

\bibitem{5PData}
David McKay.
\newblock Potassium 5{P} line data, 2009.

\bibitem{KTrans7}
Jun Jiang and J~Mitroy.
\newblock Hyperfine effects on potassium tune-out wavelengths and
  polarizabilities.
\newblock {\em Physical Review A}, 88(3):032505, 2013.

\bibitem{KTrans8}
Yashpal Singh, DK~Nandy, and BK~Sahoo.
\newblock Reexamination of nuclear quadrupole moments in 39 - 41 {K} isotopes.
\newblock {\em Physical Review A}, 86(3):032509, 2012.

\bibitem{KTrans9}
Alexandra Behrle, Marco Koschorreck, and Michael K{\"o}hl.
\newblock Isotope shift and hyperfine splitting of the 4{S} $\rightarrow$ 5{P}
  transition in potassium.
\newblock {\em Physical Review A}, 83(5):052507, 2011.

\bibitem{Boltz1}
Cun-Feng Cheng, YR~Sun, and Shui-Ming Hu.
\newblock Optical determination of the boltzmann constant.
\newblock {\em Chinese Physics B}, 24(5):3301, 2015.

\bibitem{Boltz2}
S~Mejri, Papa Sow, O~Kozlova, C~Ayari, Sean Tokunaga, C~Chardonnet,
  S~Briaudeau, B~Darqui{\'e}, F~Rohart, and C~Daussy.
\newblock Measuring the boltzmann constant by mid-infrared laser spectroscopy
  of ammonia.
\newblock {\em arXiv preprint arXiv:1506.01828}, 2015.

\end{thebibliography}

\end{document}